\documentclass[preprint,showpacs,amsmath,amssymb]{revtex4}
\usepackage{mathrsfs}

\usepackage{graphicx,color}
\usepackage{dcolumn}
\usepackage{bm}

\newcommand{\pphwa}{$pp \to HW^{\pm}\gamma+X$ }
\begin{document}


\title{Next-to-leading order QCD corrections to $HW^{\pm}\gamma$ production at the LHC}

\author{Song Mao$^a$}
\author{Wan Neng$^a$}
\author{Li Gang$^a$}\email{lig2008@mail.ustc.edu.cn}
\author{Ma Wen-Gan$^b$}
\author{Zhang Ren-You$^b$}
\author{Guo Lei$^b$}
\author{Zhou Ya-Jin$^c$}
\author{Guo Jian-You$^a$}

\affiliation{$^a$ School of Physics and Material Science, Anhui University, Hefei, Anhui 230039, P.R.China}
\affiliation{$^b$ Department of Modern Physics, University of Science and Technology of China (USTC),
                  Hefei, Anhui 230026, P.R.China}
\affiliation{$^c$ School of Physics, Shandong University, Jinan Shandong 250100, P.R. China}

\date{\today}

\begin{abstract}
The Higgs boson production associated with a $W$-boson and a photon
at the Large Hadron Collider (LHC) can be used to probe the coupling
between Higgs boson and vector gauge bosons and discover a
signature of new physics. We present the precision predictions up to
the QCD next-to-leading-order (NLO) in the standard model for this
process involving the subsequential weak decays of the final Higgs
and $W$-boson. The dependence of the leading order (LO) and the QCD
NLO corrected integrated cross sections on the
factorization/renormalization energy scale is studied. We provide
the LO and QCD NLO corrected distributions of the transverse momenta
and rapidities of final products. We find that the LO cross section
is significantly enhanced by the QCD NLO correction, and the
$K$-factor value is obviously related to the physical observables
and the phase space regions.
\end{abstract}

 \maketitle

\section{Introduction}
\par
The Higgs mechanism plays a crucial role in the standard model (SM).
The existence of the Higgs boson makes the breaking of the
electroweak (EW) symmetry and generates the masses for fundamental
particles \cite{sm,higgs}. Therefore, studying the Higgs mechanism
is one of the main goals of the Large Hadron Collider (LHC).
Recently, both ATLAS and CMS collaborations have announced the
discovery of a new boson, whose properties are compatible with that
of the SM Higgs particle, with mass of $m_H \approx 125~GeV$. Both
collaborations excluded additional Higgs-like bosons in a large mass
range of $m_H$ about $600~GeV$ \cite{higgs1, higgs2}. The
interpretation of the excesses observed in various production and
decay channels, as originating from a single spin-zero particle, was
made possible by detailed theoretical predictions for the Higgs
boson production and decay rates, see Ref.\cite{higgs3} for an
overview.

\par
After the discovery of the Higgs boson, our main task is to determine
its properties, such as spin, $CP$, and couplings. However, these
measurements require accurate theoretical predictions for both
signal and background. The process \pphwa is one of the important
processes in providing detailed information about the coupling
between Higgs boson and vector gauge bosons. This process with
subsequent decay of final state to leptons, photons and missing
energy, provides a background to new physics searches. It also
offers the possibility to study the anomalous couplings in quadrilinear vertices,
not present at tree-level in the SM such as
$H\gamma WW$, which could be directly
investigated in this process as it would cause deviations from the predicted signal.

\par
At the LHC, most of the important processes are multi-body final
state production processes. It is known that the theoretical
predictions beyond the LO for these processes with more than two
final particles are necessary in order to test the SM and search for
new physics. But the calculations for these processes involving the
NLO corrections are very intricate. In the last few years, the
phenomenological results including the NLO QCD corrections for
triple gauge boson (TGB) production processes at the LHC, such as
$pp\to WW\gamma,~ZZ\gamma,~W\gamma\gamma,~Z\gamma\gamma,~ZZZ,~ WWZ$
have been studied
\cite{Bozzi2009,Bozzi2011,Lazopoulos:2007ix,Hankele:2007sb}.
However, Higgs productions associated with di-boson at the NLO were
studied less, excepted the process $pp \to H^0 W^+ W^-$
\cite{song2009}.

\par
In this paper, we make a precision calculation for the process
\pphwa at the LHC, including the contributions of the NLO QCD
corrections. In section II we give the calculation description of
the LO cross section for the \pphwa process, and the NLO QCD
radiative contribution are presented in section III. In section IV
we present some numerical results and discussion. Finally a short
summary is given.

\vskip 5mm
\section{LO cross section for $pp \to HW^{+}\gamma+X$ }
\par
In the LO and NLO calculations we employ FeynArts 3.4
package\cite{fey} to generate Feynman diagrams and their
corresponding amplitudes. The amplitude calculations are implemented
by applying FormCalc 5.4 programs \cite{formloop}.

\par
\begin{figure}
\begin{center}
\includegraphics[width=0.6\textwidth]{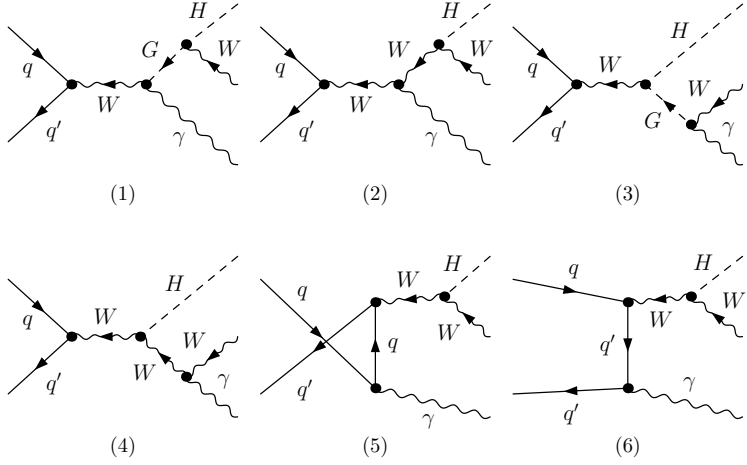}
\caption{ \label{fig1} The LO Feynman diagrams for the partonic
process $qq' \to HW^+\gamma$. }
\end{center}
\end{figure}

Due to the $CP$-conservation, the cross section for the $qq' \to
HW^{-}\gamma ~(qq' = \bar{u}d,\bar{u}s,\bar{c}d,\bar{c}s)$
subprocess in the SM should be the same as that for the
corresponding charge conjugate subprocess $qq' \to HW^{+}\gamma
~(qq' = u\bar{d},u\bar{s},c\bar{d},c\bar{s})$ at the parton level.
Therefore, we present the parton level calculations for the related
subprocess $qq' \to HW^{+}\gamma$ in this section. By taking the Cabibbo-Kobayashi-Maskawa
matrix elements $V_{td}=V_{ts}=V_{ub}=V_{cb}=0$, the LO contribution
to the cross section for the parent process $pp \to HW^+\gamma+X$
comes from the subprocesses
\begin{equation}
\label{process1} q(p_1)+ q'(p_2) \to H(p_3)+ W^+(p_4)+
\gamma(p_5),~~~~(qq' = u\bar{d},u\bar{s},c\bar{d},c\bar{s}),
\end{equation}
where $p_{1}$, $p_{2}$ and $p_{3}$, $p_{4}$, $p_{5}$ represent the
four-momenta of the incoming partons and the outgoing $H$, $W^{+}$
and photon, respectively. We use the 't Hooft-Feynman gauge in our
calculations. Since the Yukawa coupling strength is proportional to the
fermion mass and the masses of $u$-, $d$-, $s$-, and $c$-quark are
relatively small and can be neglected, we ignore the contribution
from the Feynman diagrams with internal Higgs boson line. The
Feynman diagrams for the subprocess $qq' \to HW^{+}\gamma$ at the LO
are depicted in Fig.\ref{fig1}.

\par
The expression of the LO cross section for the subprocess $qq' \to HW^{+}\gamma$
has the form as
\begin{eqnarray}
\label{LO}\hat{\sigma}^{0}_{qq'}= \frac{1}{4}\frac{1}{9}\frac{(2 \pi
)^4}{2\hat{s}}\int \sum_{spin}^{color} |{\cal M}^{LO}_{qq'}|^2
d\Omega_{3}.
\end{eqnarray}
where the factors $\frac{1}{4}$ and $\frac{1}{9}$ are due to the
averaging over the spins and colors of the initial partons,
respectively, $\hat{s}$ is the partonic center-of-mass energy
squared, and ${\cal M}^{LO}_{qq'}$ is the amplitude of all the
tree-level diagrams shown in Fig.\ref{fig1}. The summation are taken
over the spins and colors of all the relevant particles in the $qq'
\to HW^{+}\gamma$ subprocess. The integration is performed over the
three-body phase space of the final particles $H$, $W^{+}$ and
$\gamma$. The phase-space element $d\Omega_{3}$ in Eq.(\ref{LO}) is
expressed as
\begin{eqnarray}\label{PhaseSpace}
{d\Omega_{3}}=\delta^{(4)} \left( p_1+p_2-\sum_{i=3}^5 p_i \right)
\prod_{j=3}^5 \frac{d^3 \textbf{\textsl{p}}_j}{(2 \pi)^3 2 E_j}.
\end{eqnarray}
It is obvious that the LO cross section $\hat{\sigma}^{0}_{qq'}$ is
IR divergent when we integrate the Feynman amplitude squared,
$|{\cal M}^{LO}_{qq'}|^2$, over the full three-body final state
phase space. The divergence arises from the integration over the
phase space region where the final photon is soft or the photon is
radiated from one of the initial massless quarks collinearly. To
avoid these IR singularities and obtain an IR-safe result, we should
take a transverse momentum cut for final photon (see Sec.4,
Eq.(\ref{isol-1})). Then the LO total cross section for the parent
process $pp \to HW^{+}\gamma+X$ at the LHC can be expressed as
\begin{equation}
\label{integration} \sigma_{LO}= \sum_{ij=u\bar{d},u\bar{s}}^{c
\bar{d},c \bar{s}} \int_{0}^{1}dx_1 \int_{0}^{1} dx_2 \left[
G_{i/P_1}(x_1,\mu_f) G_{j/P_2}(x_2,\mu_f)+(1 \leftrightarrow
2)\right] \hat{\sigma}^{0}_{i j}(\hat{s}=x_1 x_2 s),
\end{equation}
where $G_{i/A}(x,\mu_f)$ is the parton ($i=u,c,\bar{d},\bar{s}$) distribution
function of proton (PDF) $A(=P_1,P_2)$ \cite{pdfs}, which describes the probability to
find a parton $i$ with momentum $xp_A$ in proton $A$, $s$ is defined
as the total colliding energy squared in proton-proton collision,
$\hat{s}=x_1x_2 s$, and $\mu_f$ is the factorization energy scale.

\vskip 5mm
\section{NLO QCD corrections to $pp \to HW^{+}\gamma+X$  }
\par
\begin{figure}
\begin{center}
\includegraphics[width=0.82\textwidth]{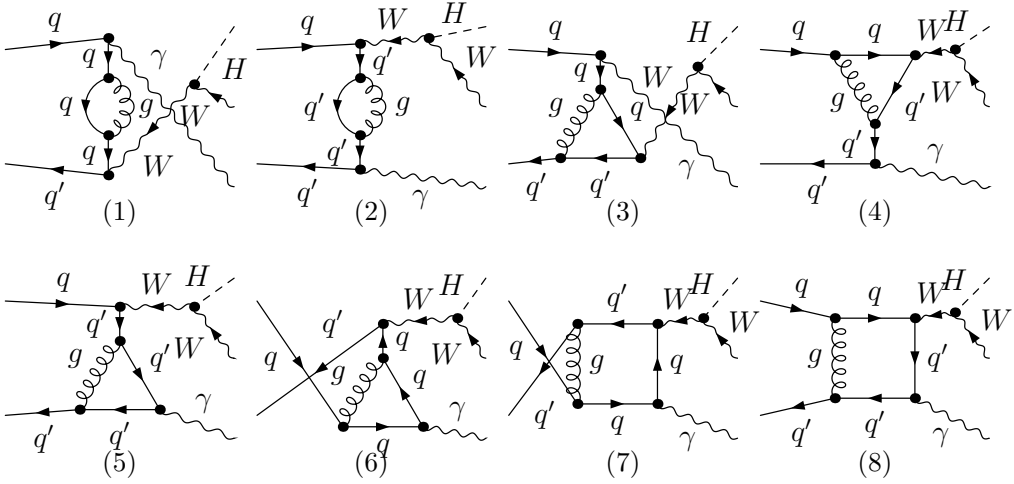}
\caption{ \label{fig2} The representative one-loop Feynman diagrams
for the partonic process $qq' \to HW^{+}\gamma$. }
\end{center}
\end{figure}

The ${\cal O}(\alpha_s)$ virtual corrections to the partonic $qq'
\to HW^{+}\gamma$ processes consist of self-energy, vertex, box and
pentagon diagrams. Part of these diagrams are presented in
Fig.\ref{fig2}. We use the definitions of tensor and scalar one-loop
integral functions in Ref.\cite{Passarino,denner2}. Using the
Passarino-Veltman (PV) method \cite{Passarino,denner3}, the tensor integrals
are expressed as a linear combination of tensor structures and coefficients, where the tensor
structures depend on the external momenta and the metric tensor, while the coefficients
depend on scalar integrals, kinematic invariants and the dimension of the integral.

\par
After the tensor integral reduction is performed, the fundamental
building blocks are one-loop scalar integrals. They may be finite or
contain both ultraviolet (UV) and infrared (IR) divergences. The UV
and IR singular scalar integrals are calculated analytically by
using dimensional regularization in $D=4-2 \epsilon$ dimensions. We
adopt the expressions in Ref.\cite{IRDV} to deal with the IR
divergences in Feynman integrals, and apply the expressions in
Refs.\cite{OneTwoThree,Four,Five} to implement the numerical
evaluations for the IR safe parts of N-point integrals. The UV
singularities of the virtual corrections are removed by introducing
a set of related counterterms.
The counterterms are defined as
\begin{eqnarray}
\psi^{0,L,R}_{q} & = & \left(1+\frac{1}{2}\delta Z_{q}^{L,R}\right)\psi^{L,R}_{q}~,
\end{eqnarray}
where $\psi^{L,R}_{q}$ denote the fields of SM
quark. The on-mass-shell scheme is adopted
to fix the wave function renormalization constant of the external
light quark field, then we obtain
\begin{eqnarray}
\delta Z^{L,R}_{q} & = & - \frac{\alpha_s (\mu_R)}{3\pi} \left[
\Delta_{UV} -
\Delta_{IR} \right]~,
\end{eqnarray}
where $\Delta_{UV}=\frac{1}{\epsilon_{UV}}-\gamma_E + \ln (4\pi)$
and $\Delta_{IR}=\frac{1}{\epsilon_{IR}}-\gamma_E + \ln (4\pi)$.

\par
After performing the renormalization procedure, the
total NLO QCD amplitude for the subprocess
$qq' \to HW^{+}\gamma$ is UV finite. Nevertheless, it still
contains soft/collinear IR singularities.
As we shall see later the soft/collinear IR singularities can
be cancelled by adding the contributions of the real
gluon/light-(anti)quark emission subprocesses, and redefining
the parton distribution functions at the NLO.

\par
\begin{figure}
\begin{center}
\includegraphics[width=0.6\textwidth]{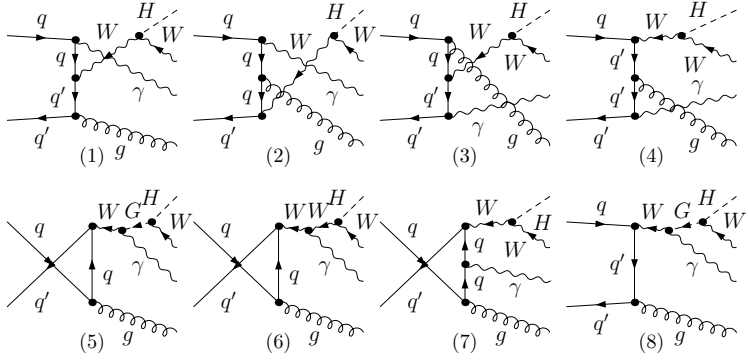}
\caption{ \label{fig3} The representative Feynman diagrams for the
real gluon emission subprocess $qq' \to HW^+\gamma g$. }
\end{center}
\end{figure}

According to the Kinoshita-Lee-Nauenberg (KLN) theorem \cite{kln} the UV
and IR singularities are exactly vanished after combining the
renormalized virtual corrections with the contributions of the real
gluon emission processes and the PDF counterterms together.
These cancelations can be verified numerically
in our numerical calculations.
The real gluon emission partonic process for the $HW^+\gamma$ can be denoted as
\begin{equation}
q(p_1)+ q' (p_2) \to H(p_3) +W^+(p_4) + \gamma(p_5) + g(p_6),~~~
(qq'~=~u\bar{d},u\bar{s},c\bar{d},c\bar{s}).
\end{equation}
The real gluon emission subprocess $qq' \to HW^+\gamma g$ (shown in
Fig.\ref{fig3}) contains both soft and collinear IR singularities.
The IR singularities can be conveniently isolated by adopting the two cutoff phase
space slicing (TCPSS) method \cite{TCPSS}, which is intuitive, simple to implement,
and relies on a minimum of process dependent information. The soft IR singularity
in the subprocess $qq' \to HW^+\gamma g$ at the LO cancels the
analogous singularity arising from the one-loop level virtual
corrections to the $qq' \to HW^+\gamma$ subprocess.

\par
In performing the calculations with the TCPSS method, we should
introduce two arbitrary small cutoffs $\delta_s$ and $\delta_c$. The
phase space of the $qq' \to HW^+\gamma g$ subprocess can be split
into two regions, $E_6 \leq \delta_s\sqrt{\hat{s}}/2$ (soft gluon
region) and $E_6 > \delta_s\sqrt{\hat{s}}/2$ (hard gluon region) by
soft cutoff $\delta_s$. The hard gluon region is separated as hard
collinear (${\rm HC}$) and hard noncollinear ($\overline{\rm HC}$)
regions by cutoff $\delta_c$. The ${\rm HC}$ region is the phase
space where $-\hat{t}_{16}$ (or $-\hat{t}_{26}$)$<\delta_c \hat{s}$
$(\hat{t}_{16}\equiv(p_1-p_6)^2$ and
$\hat{t}_{26}\equiv(p_2-p_6)^2)$. Then the cross section for the
real gluon emission subprocess can be expressed as
\begin{eqnarray}
\hat{\sigma}^R_g(q\bar{q} \to HW^+\gamma+
g)=\hat{\sigma}^S_g+\hat{\sigma}^H_g &=&
\hat{\sigma}^S_g+\hat{\sigma}^{\rm HC}_g+\hat{\sigma}^{\overline{\rm
HC}}_g.
\end{eqnarray}

\par
\begin{figure}
\begin{center}
\includegraphics[width=0.6\textwidth]{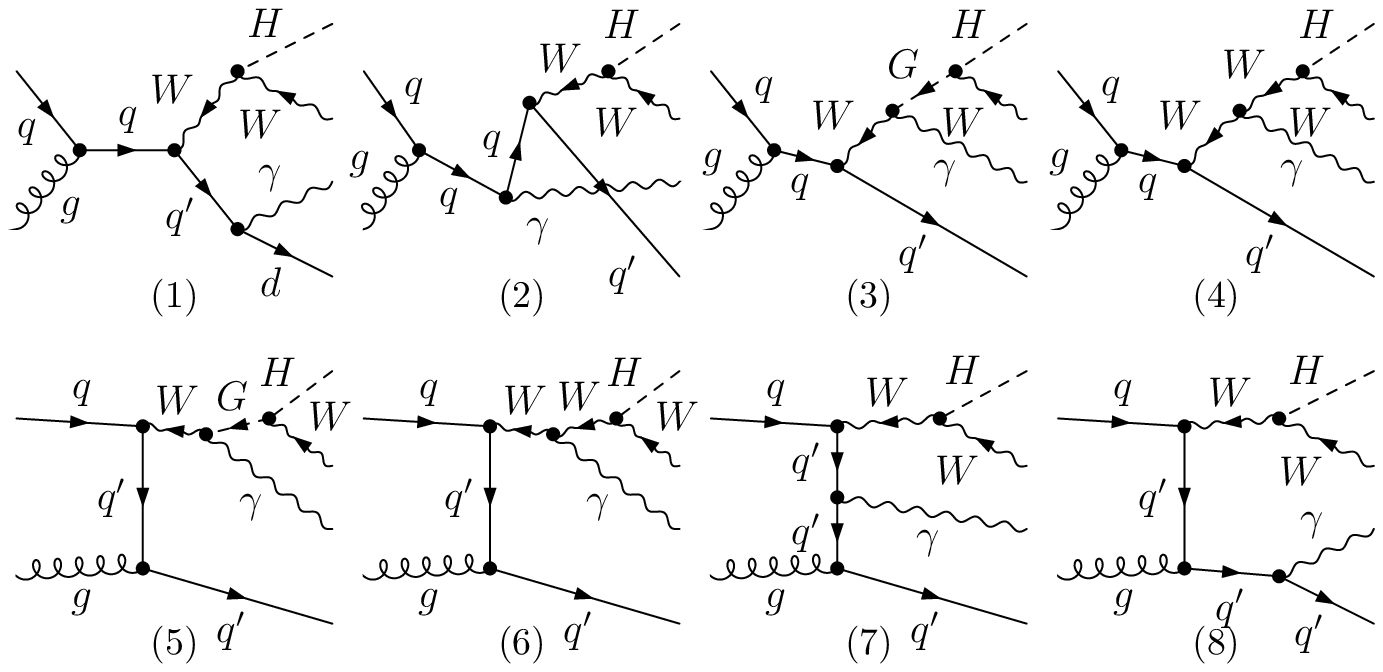}
\caption{ \label{fig4}  The representative Feynman diagrams for the
real light-quark emission subprocess $qg \to HW^+\gamma q'$. }
\end{center}
\end{figure}

Beside the real gluon emission subprocess discussed above, there is
another kind of contribution called the real light-quark emission
correction which has the same order contribution with previous real
gluon emission subprocess in perturbation theory. The corresponding
Feynman diagrams for the subprocess $qg \to HW^+\gamma q'$ at the
tree-level are shown in Fig.\ref{fig4}. We denote this subprocess as
\begin{equation}
q(p_1)+ g(p_2) \to H(p_3) +W^+(p_4) + \gamma(p_5) + q'(p_6),~~~
(qq'~=~ud,us,\bar{d}\bar{u},\bar{s}\bar{c}).
\end{equation}
It contains only the initial state collinear singularities. Using
the TCPSS method described above, we split the phase space into
collinear region and noncollinear region by introducing a cutoff
$\delta_c$. Then the cross section for the subprocess $qg \to
HW^+\gamma q'$  can be expressed as
\begin{equation}
\hat{\sigma}^R(qg \rightarrow HW^+\gamma q') = \hat{\sigma}^R_q =
\hat{\sigma}^{C}_q + \hat{\sigma}^{\overline{C}}_q.
\end{equation}

\par
The cross section $\hat{\sigma}^{\overline{C}}_{q}$ in the
noncollinear region is finite and can be evaluated in four
dimensions using Monte Carlo method, while $\hat{\sigma}^{C}_{q}$
still contains collinear singularity, which can be absorbed into the
redefinition of the PDFs at the NLO.

\par
After adding the renormalized virtual corrections and the real
gluon/light-quark emission corrections, the partonic cross sections
still contain the collinear divergences, which can be absorbed into
the redefinition of the distribution functions at the NLO. We split
the PDF counterterm, $\delta G_{i/P}(x,\mu_f)$, into two parts: the
collinear gluon emission part $\delta G_{i/P}^{(gluon)}(x,\mu_f)$
and the collinear light-quark emission part $\delta
G_{i/P}^{(quark)}(x,\mu_f)$ as
\begin{eqnarray} \label{PDFcounterterm1}
&& \delta G_{q(g)/P}(x,\mu_f) = \delta G_{q(g)/P}^{(gluon)}(x,\mu_f)
                  +\delta G_{q(g)/P}^{(quark)}(x,\mu_f),
~~~(q = u, \bar{u}, d, \bar{d}, c, \bar{c}, s, \bar{s} ).
\end{eqnarray}
We get the expressions of the counterterm parts as
\begin{eqnarray} \label{PDFcounterterm2}
&& \delta G_{q(g)/P}^{(gluon)}(x,\mu_f) =
   \frac{1}{\epsilon} \left[
                      \frac{\alpha_s}{2 \pi}
                      \frac{\Gamma(1 - \epsilon)}{\Gamma(1 - 2 \epsilon)}
                      \left( \frac{4 \pi \mu_r^2}{\mu_f^2} \right)^{\epsilon}
                      \right]
   \int_x^1 \frac{dz}{z} P_{qq(gg)}(z) G_{q(g)/P}(x/z,\mu_f), \nonumber \\
&& \delta G_{q/P}^{(quark)}(x,\mu_f) =
   \frac{1}{\epsilon} \left[
                      \frac{\alpha_s}{2 \pi}
                      \frac{\Gamma(1 - \epsilon)}{\Gamma(1 - 2 \epsilon)}
                      \left( \frac{4 \pi \mu_r^2}{\mu_f^2} \right)^{\epsilon}
                      \right]
   \int_x^1 \frac{dz}{z} P_{qg}(z) G_{g/P}(x/z,\mu_f),  \nonumber \\
&& \delta G_{g/P}^{(quark)}(x,\mu_f) =
   \frac{1}{\epsilon} \left[
                      \frac{\alpha_s}{2 \pi}
                      \frac{\Gamma(1 - \epsilon)}{\Gamma(1 - 2 \epsilon)}
                      \left( \frac{4 \pi \mu_r^2}{\mu_f^2} \right)^{\epsilon}
                      \right]
   \sum_{q=u,\bar{u},d,\bar{d},}^{c, \bar {c}, s, \bar {s}}
   \int_x^1 \frac{dz}{z} P_{gq}(z) G_{q/P}(x/z,\mu_f).~~~
\end{eqnarray}
More details about the explicit expressions for the splitting functions
$P_{ij}(z) (ij=qq,qg,gq,gg)$ are available in Ref.\cite{TCPSS}.

\par
Finally, we have eliminated all the UV and IR singularities by
performing the renormalization procedure and adding all the NLO QCD
correction components, and we get the finite NLO QCD corrected
integrated cross section for the $pp \to HW^{+} \gamma +X$ process as
\begin{eqnarray}\label{TotalCorr}
\sigma_{NLO}&=&\sigma_{LO}+\Delta
\sigma_{NLO}= \sigma_{LO}+\Delta\sigma^{(3)}+\Delta\sigma^{(4)}.
\end{eqnarray}
The three-body term $\Delta \sigma^{(3)}$ includes the one-loop
corrections to the  process $pp \to HW^{+} \gamma +X$ and the tree-level
contributions in the soft and hard collinear regions for the real
gluon/light-(anti)quark emission processes, while the four-body
term $\Delta \sigma^{(4)}$ contains the cross sections for the real
gluon/light-(anti)quark emission processes over the hard noncollinear region.

\vskip 5mm
\section{Numerical results and discussion}
\par
In this section we present and discuss the numerical results for the
\pphwa process at both the LO and the QCD NLO. We adopt the CTEQ6L1
and CTEQ6M parton densities with five flavors in the LO and NLO
calculations, respectively.  The strong coupling
constant is determined by taking one-loop and two-loop running
$\alpha_{s}(\mu)$ for the LO and the NLO calculations separately,
and setting the QCD parameter $\Lambda_5^{LO} = 165~MeV$ for the
CTEQ6L1 at the LO and $\Lambda_5^{\overline{MS}} = 226~MeV$ for the
CTEQ6M at the NLO. For simplicity we define the factorization scale
and the renormalization scale being equal, and take $\mu\equiv\mu_f
= \mu_r = (m_H+ m_W)/2$ by default unless stated otherwise. We adopt
$m_u=m_d=m_c=m_s=0$ and employ the following numerical values for
the relevant input parameters: \cite{hdata}
\begin{equation}
\begin{array}{lll}  \label{input1}
\alpha(m_Z)^{-1}=127.918, &m_W=80.398~{\rm GeV},    &m_Z=91.1876~{\rm GeV}. \\
\end{array}
\end{equation}
The CKM matrix elements are fixed as
\begin{equation}
\begin{array}{lll}  \label{input2}
V_{CKM}=\left(                 
\begin{array}{ccc}   
 V_{ud} & V_{us} & V_{ub}\\
 V_{cd} & V_{cs} & V_{cb}\\
 V_{td} & V_{ts} & V_{tb}\\
\end{array}\right)=
\left(                 
\begin{array}{ccc}   
 0.97418 & 0.22577 & 0\\
 -0.22577 & 0.97418 & 0\\
 0 & 0 & 1\\
\end{array}\right).
\end{array}
\end{equation}

\par
In order to get rid of the IR singularity from the electroweak
sector at the LO, we take the transverse momentum cut on final
photon as
\begin{eqnarray} \label{isol-1}
p_{T_{\gamma}}>20~GeV .
\end{eqnarray}
To remove the collinear singularity between the photon and a
massless parton $i$ at the NLO calculation, we adopt the selection
criterion provided in Ref.\cite{Frixione98}. There we accept the
$HW^{\pm}\gamma$ production only if
\begin{eqnarray} \label{isol-2}
p_{T_i}\leq p_{T_{\gamma}}\frac{1-\cos R_{\gamma i}}{1-\cos\delta_0}
~~~~~or ~~~~~R_{\gamma i}> \delta_0,
\end{eqnarray}
where $\delta_0$ is a fixed separation parameter which we set it to
be $0.7$. The condition of Eq.(\ref{isol-2}) allows final state
partons arbitrarily close to the photon axis as long as they are
soft enough. In this way, we can preserve the full QCD singularity,
which cancels against the virtual part, but it does not introduce
divergences from the interaction between photon and massless quark.

\par
\begin{figure}[htbp]
\begin{center}
\includegraphics[width=0.45\textwidth]{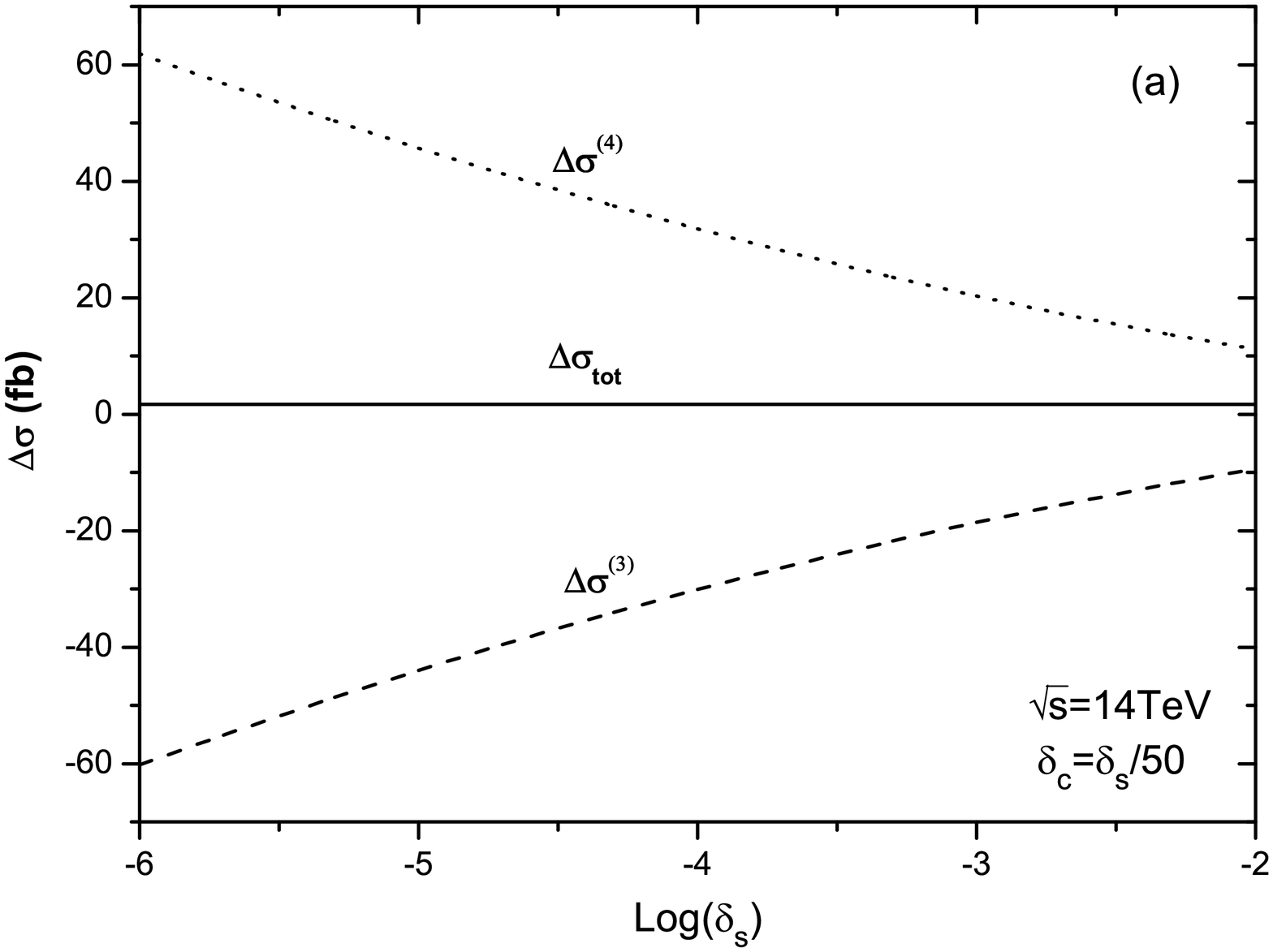}
\includegraphics[width=0.45\textwidth]{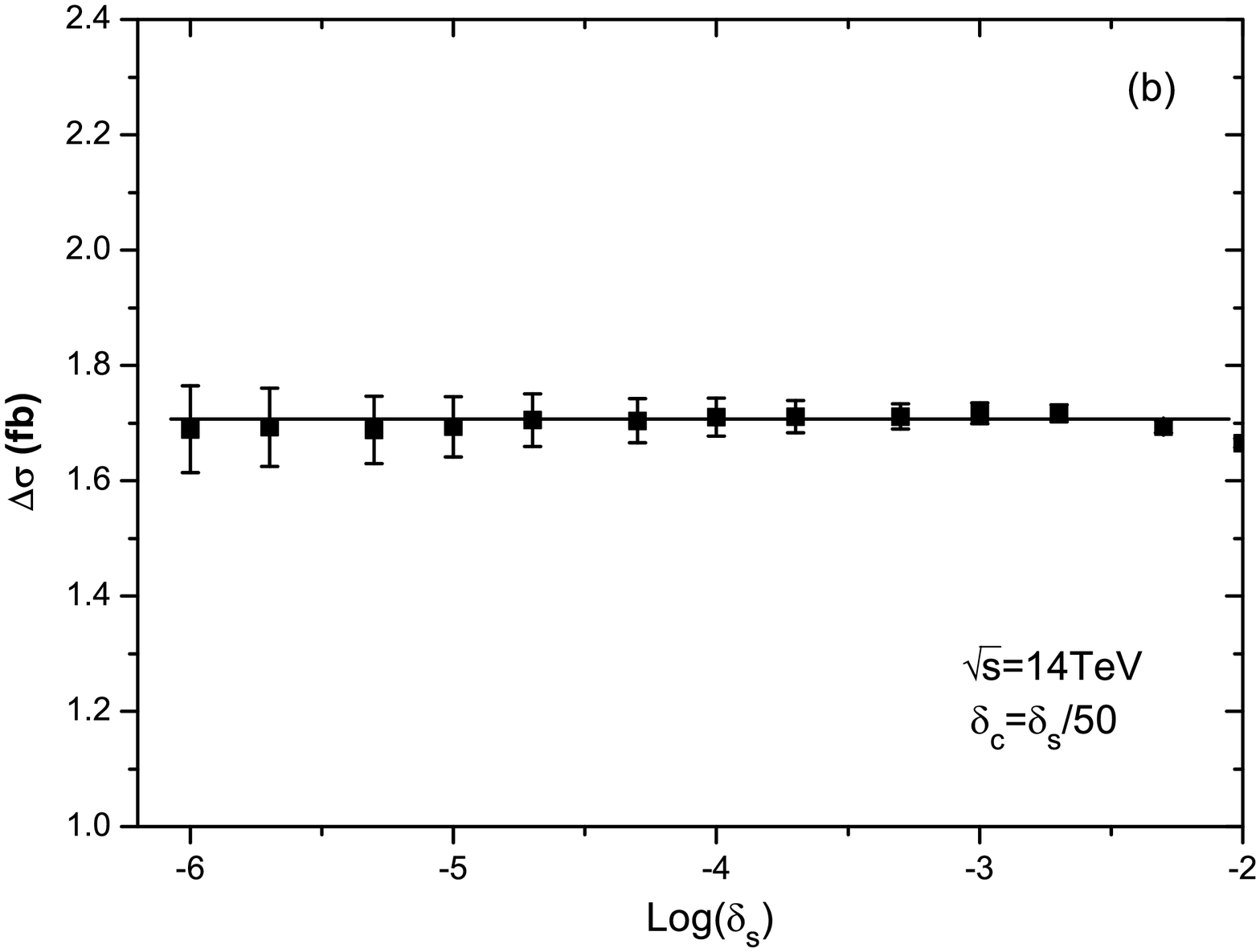}
\hspace{0in}%
\caption{\label{fig5} (a) The dependence of the NLO QCD corrections
to the \pphwa process on the cutoffs
$\delta_s$ and $\delta_c$ at the $\sqrt{s}=14~{\rm TeV}$ LHC, where we
take $\delta_c=\delta_s/50$ and
$\mu=\mu_0=(m_H+m_W)/2$. (b) The amplified curve for
$\Delta\sigma_{tot}$ in Fig.\ref{fig5}(a)}
\end{center}
\end{figure}

In Figs.\ref{fig5}(a,b), the NLO total cross section is plotted
against $\delta_s$ and $\delta_c$ to the \pphwa process at the LHC.
The amplified curve for the total correction $\Delta\sigma_{NLO}$ in
Fig.\ref{fig5}(a) is demonstrated in Fig.\ref{fig5}(b) together with
calculation errors. Using the TCPSS method, one required $\delta_c\ll\delta_s$.
For many calculations it has been found that choosing $\delta_c$ to be 50 - 100 times
smaller than $\delta_s$ is sufficient for answers accurate to a few percent.
Here, we take $\delta_c=\delta_s/50$ and
$\mu=\mu_0=(m_H+m_W)/2$.
For the NLO corrections, the virtual and real
radiation corrections depend on $\delta_s$ and $\delta_c$,
separately. However, when all pieces are added together, the
dependence on $\delta_s$ and $\delta_c$ is canceled as long as
sufficiently small values of $\delta_s$ and $\delta_c$ are chosen.

\par
\begin{figure}
\begin{center}
\includegraphics[width=0.6\textwidth]{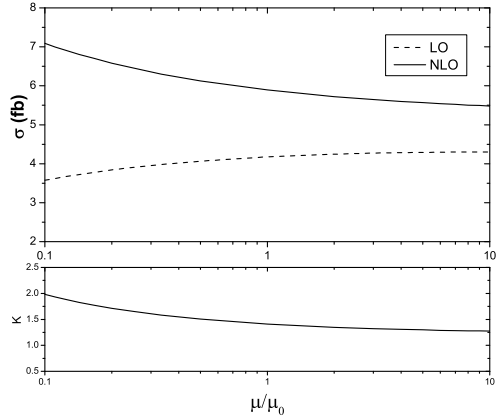}
\caption{ \label{fig6}  The dependence of the LO, NLO QCD corrected
total cross sections and the corresponding K-factor ($K(\mu)\equiv
\sigma_{NLO}(\mu)/\sigma_{LO}(\mu)$) for the process \pphwa on the
factorization/renormalization scale($\mu/\mu_0$). Here we assume
$\mu=\mu_r=\mu_f$ and define $\mu_0 = (m_H + m_W)/2$. }
\end{center}
\end{figure}

In Fig.\ref{fig6} we illustrate the renormalization and
factorization scale dependence of the LO, NLO QCD corrected total
cross sections and the corresponding $K$-factor ($K(\mu)\equiv
\sigma_{NLO}(\mu)/\sigma_{LO}(\mu)$) for the process \pphwa. We
assume $\mu=\mu_r=\mu_f$ and define $\mu_0 = (m_H + m_W)/2$. From
this figure, we can see that the LO and NLO QCD corrected total
cross section are $4.18 fb$ and $5.89 fb$ respectively, the
corresponding K-factor is $1.41$ at $\mu_r=\mu_f=\mu_0$. When the
scale $\mu$ running from $0.5 \mu_0$ to $2 \mu_0$, the related
theoretical uncertainty amounts to $^{+1.63}_{-2.61}\%$ at the LO
and to $^{+3.90}_{-2.88}\%$ at the NLO. It demonstrates that the LO
curve underestimates the energy scale dependence. That is because
the LO partonic processes for the \pphwa processes are pure
electroweak channels where the $\mu_r$ dependence is invisible at
the LO, the energy scale dependence is the consequence of the parton
distribution functions being related to the factorization scale
($\mu_f$).

\par
\begin{figure}
\centering
\includegraphics[width=0.49\textwidth]{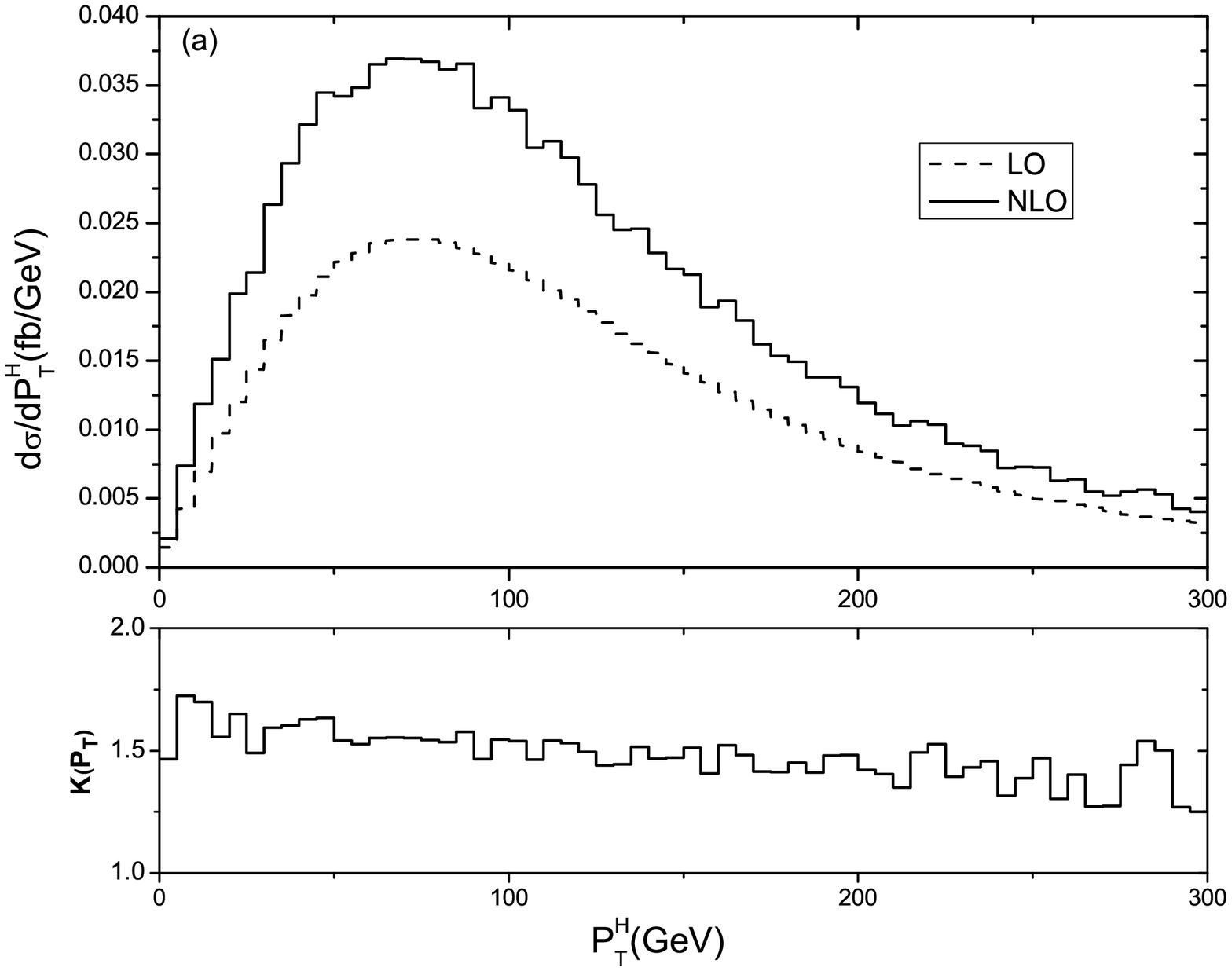}
\includegraphics[width=0.49\textwidth]{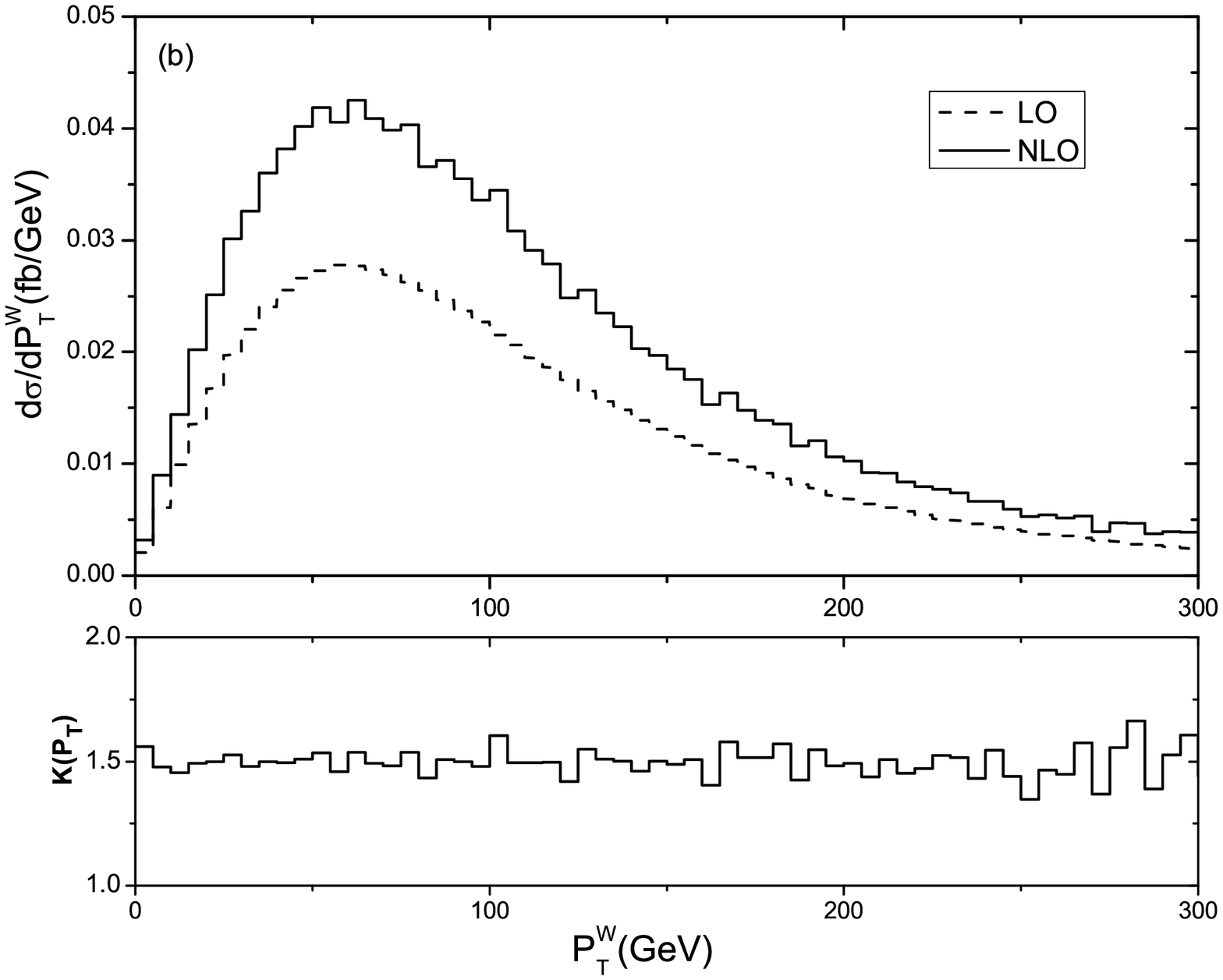}
\includegraphics[width=0.49\textwidth]{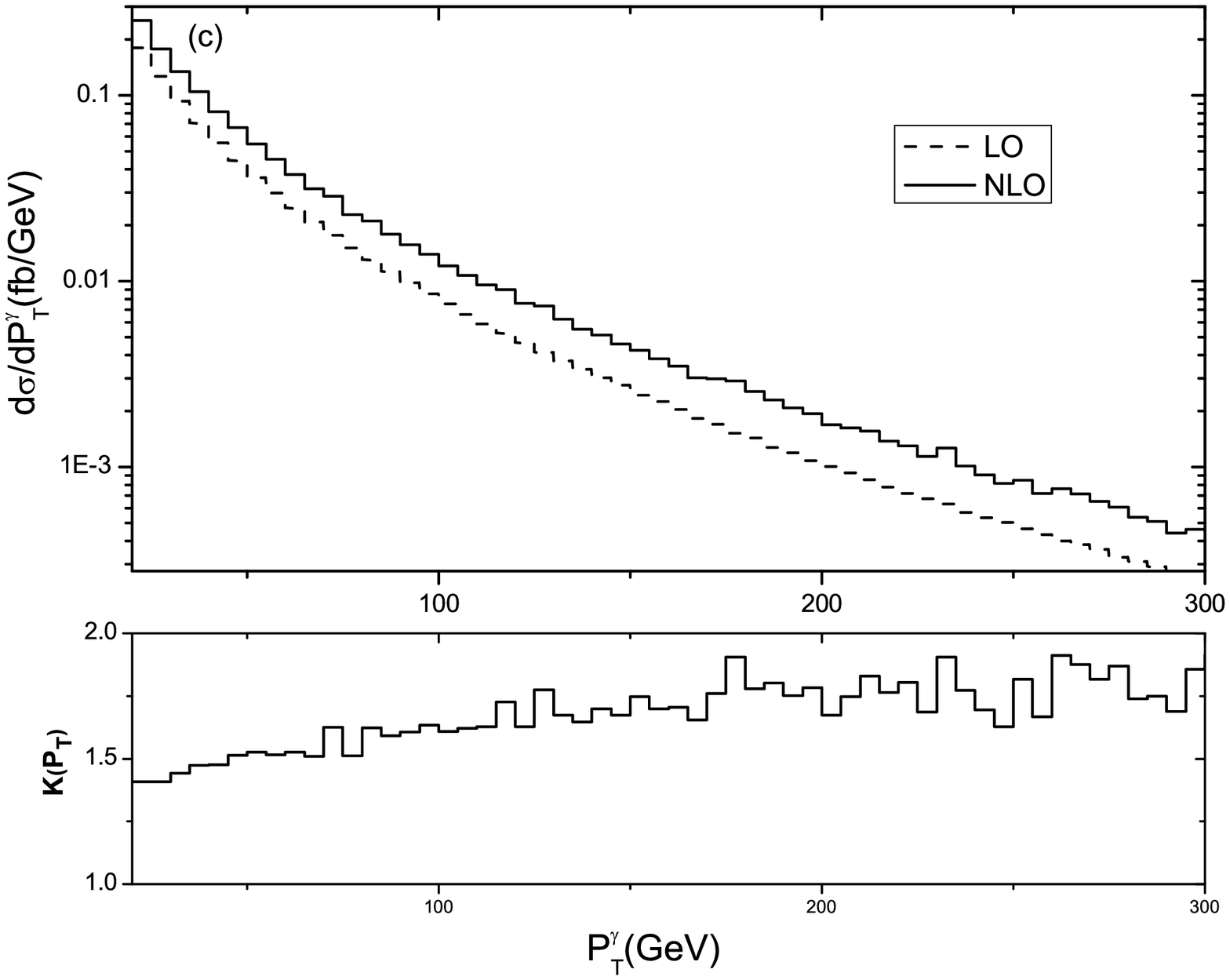}
\caption{\label{fig7} The LO and QCD NLO distributions of the
transverse momenta of final particles and the corresponding
K-factors for the \pphwa process at the LHC. (a) for $H$-boson, (b)
for $W$-boson, (c) for photon. }
\end{figure}

In Figs.\ref{fig7}(a,b,c) we depict the LO and NLO QCD corrected
differential cross sections of the transverse momenta for the final
produced $H$-, $W^\pm$-boson and photon in the process \pphwa at the
$14~TeV$ LHC. The differential cross sections of the $p_T$ for
$H$-boson at the LO and QCD NLO, i.e.,
$\frac{d\sigma_{LO}}{dp_T^{H}}$ and
$\frac{d\sigma_{NLO}}{dp_T^{H}}$, are depicted in Fig.\ref{fig7}(a),
the distributions of $\frac{d\sigma_{LO}}{dp_T^{W}}$ and
$\frac{d\sigma_{NLO}}{dp_T^{W}}$ for $W$-boson are plotted in
Fig.\ref{fig7}(b) and the distributions of
$\frac{d\sigma_{LO}}{dp_T^{\gamma}}$ and
$\frac{d\sigma_{NLO}}{dp_T^{\gamma}}$ for photon are plotted in
Fig.\ref{fig7}(c) separately. In Figs.\ref{fig7}(a) and (b), there
exist peaks for the curves of $\frac{d\sigma}{dp_T^{H}}$ and
$\frac{d\sigma}{dp_T^{W}}$ at the LO and NLO QCD corrections. The
peaks are located at the positions around $p_T^H \sim 70~GeV$ for
Higgs boson and $p_T^W\sim 60~GeV$ for $W$ boson. In
Fig.\ref{fig7}(c), we find that the differential cross section of
the photon decreases fast as the increment of the transverse momentum
of the photon. We can see from Figs.\ref{fig7}(a-c) that all the
differential cross sections at the LO for $H$-, $W$-boson and photon
($d\sigma_{LO}/dp_T^{W}$, $d\sigma_{LO}/dp_T^{H}$,
$d\sigma_{LO}/dp_T^{\gamma}$)) are significantly enhanced by the NLO
QCD corrections.

\par
\begin{figure}[htbp]
\begin{center}
\includegraphics[width=0.45\textwidth]{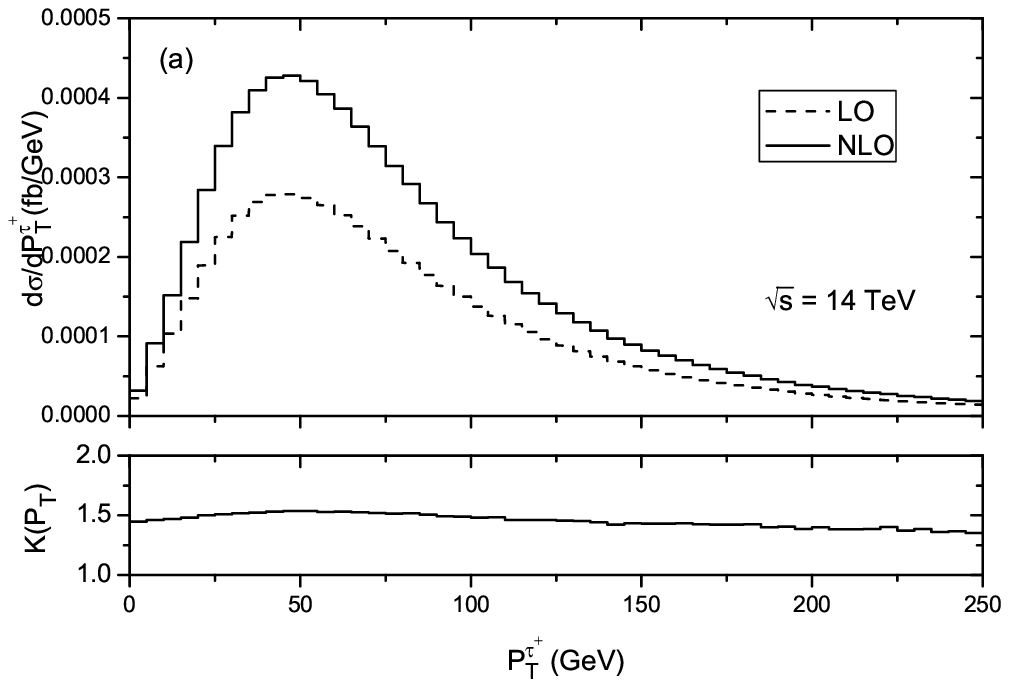}
\includegraphics[width=0.45\textwidth]{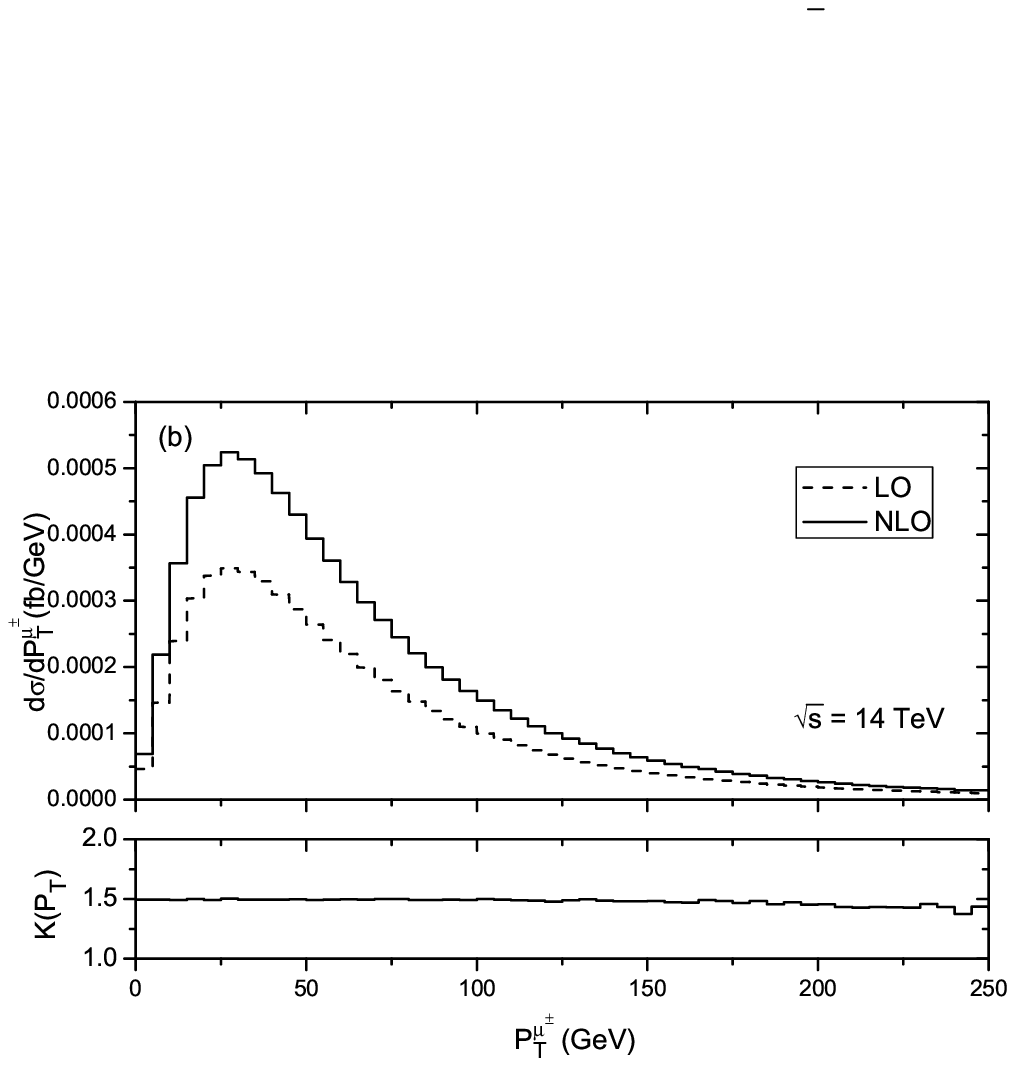}
\\~~  \\
\includegraphics[width=0.45\textwidth]{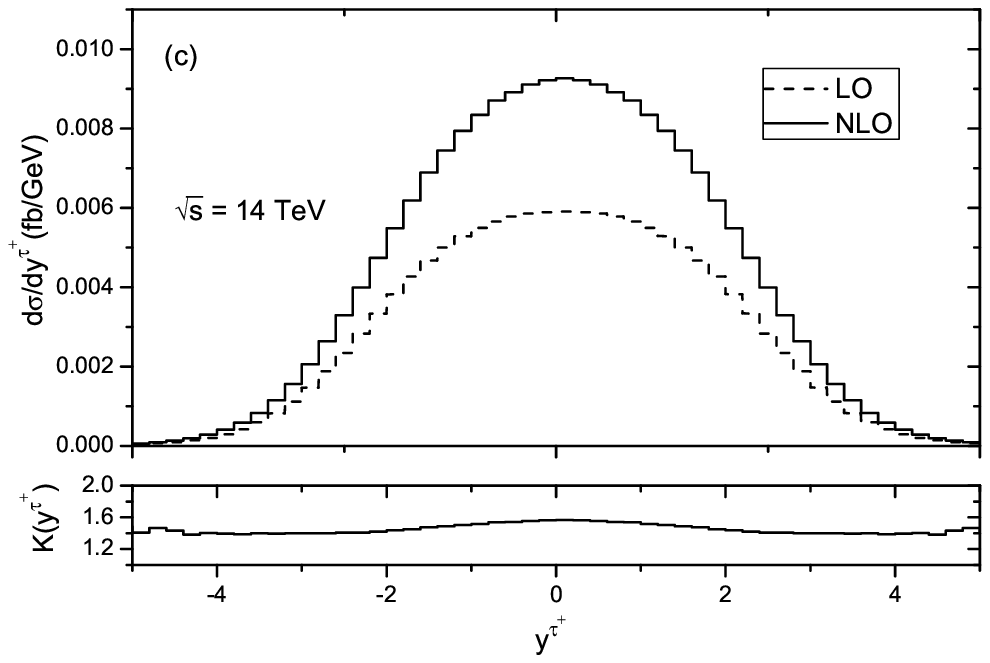}
\includegraphics[width=0.45\textwidth]{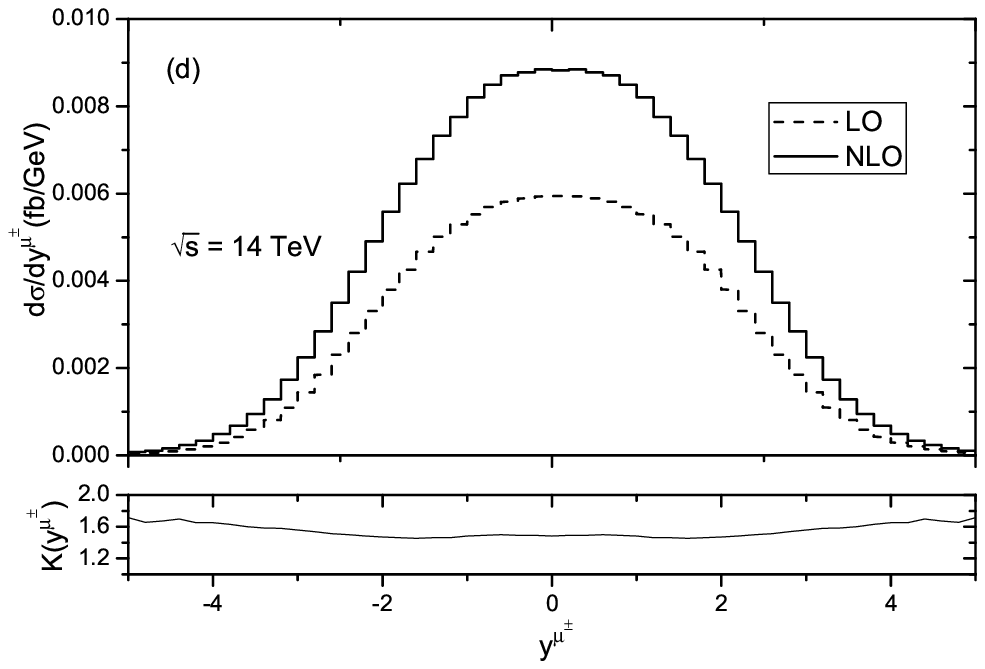}
\caption{\label{fig8} The LO and NLO QCD corrected distributions of
the transverse momenta and rapidity distributions of the final
$\tau^+$- and $\mu^{\pm}$-leptons and the corresponding $K$-factors
for the $pp \to HW^{\pm}\gamma+X \to
\tau^+\tau^-\mu^{\pm}\nu_{\mu}\gamma+X$ processes at the
$\sqrt{s}=14~TeV$ LHC.  (a) $p_T^{\tau^+}$ distributions for final
lepton $\tau^+$, (b) $p_T^{\mu^{\pm}}$ distributions for final
lepton $\mu^{\pm}$, (c) $y_T^{\tau^+}$ distributions for final
lepton $\tau^+$, (d) $y^{\mu^{\pm}}$ distributions for final lepton
$\mu^{\pm}$. }
\end{center}
\end{figure}

As we know, $\gamma$ can be detected directly in experiment, but $H$
and $W^\pm$ boson are unstable and detected by the signals of their
weak decay products. We investigate the kinematic distributions of
final products after the subsequent decays of Higgs boson and $W$
gauge boson (i.e., $H \to \tau^+\tau^-$ and $W^\pm \to
\mu^\pm\nu_{\mu}$). We employ the SM leptonic decay branch ratios of
$H$ and $W$ bosons in further numerical calculations, i.e., $Br(H^0
\to \tau^+\tau^-)=6.5\%$ and $Br(W^- \to \mu^-\bar{\nu}_\mu)=
10.57\%$ \cite{hdata}. The $HW^+\gamma$ production at the LHC
including their subsequent decays can be written as
\begin{eqnarray}\label{channel}
pp \to HW^{\pm}\gamma \to
\tau^+\tau^-\mu^\pm\stackrel{(-)}{\nu_{\mu}}\gamma +X.
\end{eqnarray}
Then a signal event of $HW^{\pm}\gamma$ production is detected at
the LHC as $\tau$-pair and one charged lepton $\mu^\pm$ plus missing
energy ($\stackrel{(-)}{\nu_{\mu}}$). In Figs.\ref{fig8}(a) and (b)
we present the LO, NLO QCD corrected distributions of the transverse
momentum of $\tau^+$ and $\mu^{\pm}$, and the corresponding
$K$-factors at the $\sqrt{s}=14~TeV$ LHC, separately. From
Figs.\ref{fig8}(a,b) we can see that the QCD corrections always
enhance the LO differential cross section
$d\sigma_{LO}/dp_T^{\tau}$, and both the LO and QCD NLO corrected
distributions of final $\tau(\mu)$ lepton at the future LHC have
their peaks at the position of $p_T^{\tau^+}\sim 50~GeV$
($p_T^{\mu^{\pm}}\sim 25~GeV$), and the $K(p_T)$-factor value can be
beyond $1.50$. Figs.\ref{fig8}(c) and (d) are for the rapidity
distributions of $\tau^+$- and $\mu^{\pm}$-leptons, respectively. We
can see from all these four figures that the NLO QCD corrections do
not make shape change in the transverse momentum and rapidity
distributions, while the NLO QCD corrections enhance the LO
differential cross sections significantly in all the plotted
kinematic regions, and the $K(y_T)$-factors for lepton $\tau^+$ and
$\mu^{\pm}$ can go beyond the values of $1.50$ and $1.70$,
respectively.

\vskip 5mm
\section{Summary}
\par
In this paper we investigate the phenomenological effects induced by
the NLO QCD corrections in the Higgs boson production associated
with a $W$-boson and a photon at the LHC. We present the dependence
of the LO and the NLO QCD corrected cross sections on the
fctorization/renormalization energy scale, and it shows that the
scale dependence of the integrated cross section is underestimated
by the LO result. We present the LO and the QCD NLO distributions of
the transverse momenta and rapidities of final particles. We find
that the NLO QCD radiative corrections obviously modify the LO
kinematic distributions, and values of $K$-factor are obviously
related to the phase space regions and the kinematic observables. It
shows that we should consider the NLO QCD corrections in precision
experimental data analyse in studying this process.

\section{Acknowledgments}
This work was supported in part by the National Natural Science Foundation of China (No.11205003,
No.11275190, No.11075150, No.11175001,No.11105083), the Key Research Foundation of Education
Ministry of Anhui Province of China (No.KJ2012A021), the Youth Foundation of Anhui Province(No.1308085QA07),
and financed by the 211 Project of Anhui University (No.02303319).


\end{document}